\documentclass[12pt]{article}
\usepackage[cp1251]{inputenc}
\usepackage{amsmath,amsthm,amssymb}
\usepackage{latexsym}
\usepackage{amsfonts}
\usepackage{xcolor}
\usepackage{pstricks-add,xcolor}

 \numberwithin{equation}{section}

 \allowdisplaybreaks

\allowdisplaybreaks

\newtheorem{Pa}{Paper}[section]
\newtheorem{Tm}[Pa]{{\bf Theorem}}
\newtheorem{La}[Pa]{{\bf Lemma}}

\begin{document}
\begin{center}
{\bf \Large Fundamental theorem of asset pricing: a strengthened version and  $p$-summable markets  \\
\ \\
 }

\medskip
\bigskip
A. V.  Lebedev, \ \  P. P. Zabreiko

\end{center}
\bigskip
\begin{abstract}

In the  article a strenthened version  of the 'Fundamental Theorem of asset Pricing' for one-period market model is proven.
The principal role in this result  play total and nonanihilating cones.
\end{abstract}
\medbreak


\medbreak
\vspace{5mm}

\tableofcontents

\section*{Introduction}\label{s.1}

One of the corner stones of the theory of Mathematical Finance is the so-called 'Fundamental Theorem of asset Pricing' (in fact there is a series of results under this name). The Fundamental Theorem of asset Pricing  links arbitrage free markets (i.e markets that do not admit riskless claims yielding profit with strictly positive probability; the accurate  definition will
be given below in Section~\ref{s.0}) with existence of martingales generated by measures that are equivalent to the initial one.
One should mention quite a number of researchers who contributed to the theme. Among them are F.~Black, M.~Scholes, R.~Merton, J.~Harrison, S.~Pliska,  S.~Ross, D.M.~Kreps, R.~Dalang, A.~Morton, W.~Willinger,  D.~Kramkov, J.~Jacod, A.N.~Shiryaev,  F.~Delbaen, W.~Schachermayer and many others.
We cannot give a full account of  sources and names related to the subject  and refer, for example,  to  \cite{Del-Schach},   \cite{Schach}, and \cite {Pasc} and the sources quoted therein.

In the article \cite{ZL} the authors described  geometry of Banach  structures forming mathematical base of the
'Fundamental Theorem of asset Pricing' type phenomena for the one-period financial market model. One of the main results of \cite{ZL} is recalled in  Theorem~\ref{t.'4''} of the present article.
There are two  principal assumptions in this theorem: the first one is  nonemptiness of the interior of the cone $\overset{\circ}{K^*} \neq \emptyset$, where $K$ is the cone of arbitrage possibilities (profit cone), and the second one is reflexivness of the subspace  $L=L^{**}$, where   $L$ as the subspace of financial market strategies. Both these conditions are trivially satisfied
in the standard situation  usually considered in the 'Fundamental Theorem of asset Pricing' (cf. Theorems~\ref{t.4} and~\ref{t.4'} of the present article), namely, $K=L_{1+}$ is the cone of nonnegative functions in  $L_1(\Omega, P)$ and $L$ is a finite dimensional subspace.
As is shown in \cite{ZL}, Theorem~4.1 once one relaxes essentially conditions on $K$ (for example, do not presupposes satisfaction of the property $\overset{\circ}{K^*} \neq \emptyset$) martingality somehow disappears from arbitrage freeness criterium: in a general situation there is no condition like condition 2) in Theorem~\ref{t.'4''} and only a certain analog of condition 3) takes place. Moreover,  arbitrage freeness criterium  itself reduces to the classical theorem on bipolar (see the proof of Theorem~4.1, \cite{ZL}).  One of the goals of the present article is the analysis of  relaxation of conditions on  $K$ without loosing the martingality nature of arbitrage free markets. We will find out that the principal condition under which the martingality
nature of arbitrage free markets is still alive is  existence of total and nonanihilating cones in $K^*$. Such cones can radically reduce the 'search region' for martingale measures.
As is shown by concrete examples this region can be essentially smaller than $\overset{\circ}{K^*}$ and in fact often exists even when $\overset{\circ}{K^*} = \emptyset$. In addition in a number of situations this observation makes it possible to strengthen  Theorems~\ref{t.4},~\ref{t.4'}, and~\ref{t.'4''}.

The article is organized as follows. In the starting Section~\ref{s.0} we recall a one-period market model and the corresponding Fundamental Theorem of asset Pricing. Then we give  a geometric reformulation of this theorem  (Theorem~\ref{t.4}) and formulate its  refined version (Theorem~\ref{t.4'}) which, in fact, is a  result from \cite{ZL}. Here we also give a general description of arbitrage free markets geometry (Theorem~\ref{t.'4''}; \cite{ZL}, Theorem~2.4).
The next Section~\ref{s.2'} presents the principal results of the  article. We introduce  total and nonanihilating cones and give examples of these objects.
The main result (Theorem~\ref{str-principal-t})  is  a strengthened version of the  Fundamental Theorem of asset Pricing  exploiting objects in question.
As a particular corollary of this result and examples considered we obtain the corresponding Fundamental Theorem of asset Pricing for $p$-summable markets (Theorem~\ref{L-p}).
We finish the article with a  comment on a finite dimentional situation where we find out that the set of objects in question reduces to a single element (Theorems~\ref{t.finite-dim}~ and~\ref{t.finite}).

\section{Arbitrage free markets geometry}\label{s.0}

In this section we recall  the Fundamental Theorem of asset Pricing for one-period market model and discuss its geometric nature.  It will be the starting point of our further analysis.

A one-period  market model is given in the following way. Let us denote by
\label{p.1}$\overline{\pi} := (\pi^0,\,\pi ):= (\pi^0,\pi^1, \ldots , \pi^d ) \in \mathbb{R}^{d+1}_+$ the (initial, known) price system at moment  $t_0$. By
$\overline{S}:= (S^0, \,S):=(S^0, S^1, \ldots , S^d)$ we denote the   price system at moment  $t_1$, that is a family of  {nonnegative} random variables on a probability space  $(\Omega, \cal{F}, P)$ \ (where $(\Omega, \cal{F}, P)$ is the space of  (possible) \emph{scenario}). It is assumed that all the random variables under consideration are summable, that is   $S^i \in L_1(\Omega, P), \ i=\overline{0,d}$. In addition the variable
   $S^0$ is assumed to be a riskless bond, that is  it is not random
\begin{equation}
 \label{e.0}
S^0:\equiv (1+r)\,\pi^0,
\end{equation}
  where $r$ is interpreted as a  bank interest rate (for purely mathematical reasons one can assume that  $r>-1$). In what follows we presuppose that
 \begin{equation}
 \label{e.1}
 \pi^0 =1,
 \end{equation}
that is the price  $\pi^0$ is normalized. Therefore
\begin{equation}
 \label{e.2}
S^0:\equiv (1+r);
\end{equation}
  A (starting) investment  \emph{portfolio} is a vector  \  $\overline{\xi}:= (\xi^0, \xi ):=(\xi^0, \xi^1, \ldots, \xi^d )\in \mathbb{R}^{d+1}$, where the values   $\xi^i$ can be negative.

\noindent The price of buying the portfolio (at moment $t_0$) is equal to
\begin{equation}
 \label{e.3}
\overline{\xi}\cdot\overline{\pi} := \sum_{i=0}^d\xi^i \, \pi^i\, .
\end{equation}
And the value of portfolio (at moment $t_1$) is the random variable
\begin{equation}
 \label{e.4}
\overline{\xi}\cdot\overline{S} = \sum_{i=0}^d\xi^i \, S^i (\omega) =  \xi^0 \, (1+r)+ {\xi}\cdot {S}\,.
\end{equation}
An \emph{arbitrage opportunity} is a portfolio  $\overline{\xi}\in \mathbb{R}^{d+1}$, such that
$$
\overline{\xi}\cdot\overline{\pi}\le 0, \ \ \text{but} \ \ \left\{ \overline{\xi}\cdot\overline{S} \ge_{\text{a.e.}}0 \ \ \text{and} \ \ P(\overline{\xi}\cdot\overline{S}> 0) \, >0\right\}.
$$
If the market is arbitrage free  (i.e. there are no portfolio satisfying the relation written above) it is reasonable to consider it as being just.

It is convenient to express the arbitrage freeness conditions in terms of the so-called  {discounted net gains}.
Recall that  \emph{discounted net gains} (at moment  $t_1$) are the random variables given by
\begin{equation}
\label{e.1'}
Y^i := \frac{S^i}{1+r} -\pi^i, \ \ \ i=1,\ldots,d\,.
\end{equation}
 Let us denote by  $Y$ the vector of discounted net gains
$Y:= (Y^1, \ldots, Y^d)$.

\noindent By  \eqref{e.0} we have   $Y^0 = \frac{S^0}{1+r} -\pi^0 =0$ and therefore  $Y^0$ does not play any role.
\begin{La}
\label{t.1}
{\rm [\cite{Fol-Sch}, Lemma~1.3 and condition (1.3)]} \ The following conditions are equivalent:

{\rm 1)} market is arbitrage free;

{\rm 2)} if  $\xi \in \mathbb{R}^d$ satisfies  $\xi\cdot Y \ge_{\text{a.e.}} 0$, then  $\xi\cdot Y =_{\text{a.e.}} 0$.
\end{La}
\smallskip

This lemma has clear geometric interpretation.

\noindent Let
\begin{equation}
\label{e.8}
 L := \{\xi\cdot Y, \ \xi \in  \mathbb{R}^d\} = \left\{ \sum_{i=1}^d \xi_i\, Y^i, \ \ (\xi_1, \ldots , \xi_d)\in \mathbb{R}^d\right\}
\end{equation}
be the subspace generated by the vectors  (functions) $Y^i, \ i=1,\ldots,d$. By  $L_{1+}$ we denote the cone of nonnegative functions
\begin{equation}
\label{e.0'}
L_{1+} := \{f\in L_1(\Omega, P), \ f\ge 0\}.
\end{equation}
 Lemma~\ref{t.1} means that
\begin{equation}
\label{e.5}
\text{\emph{a market is arbitrage free}} \ \Leftrightarrow \ \ L\cap L_{1+} = \{0\}.
\end{equation}
The foregoing observations make it natural to consider $L_{1+}$ as the cone of \emph{arbitrage possibilities}   (\emph{profit cone}) and consider $L$ as the subspace of \emph{financial market strategies}.
\smallskip

Evidently it is important to obtain description of (geometric, algebraic and etc.) conditions under which the equality   $L\cap L_{1+} = \{0\}$ takes place.  The most known market arbitrage freeness condition in financial mathematics is the \emph{fundamental theorem of asset pricing}. It sounds as follows (see, for example, \cite{Fol-Sch}, Theorem~1.6):
\smallskip

\noindent\emph{a market is arbitrage free  \ $\Leftrightarrow$ \ \ there exists a martingale measure  $P^*$ which is equivalent to the initial measure $P$ and has a bounded Radon-Nikodim derivative~$\frac{d\,P^*}{d\,P}$.}
\medskip

In this article we prove  a certain strengthened version of this result (Theorem~\ref{str-principal-t}). To implement this we need  a geometric reformulation of the above mentioned fundamental theorem of asset pricing.  Hereafter we present it.
\smallskip

Let us consider a Banach space  $L_1(\Omega, P)$. As usually, elements of this space are equivalence classes of integrable functions, where the equivalence of two functions is given by their equality almost everywhere; and the norm is given by the integral.    Thus, all the equalities and inequalities are understood as 'almost everywhere'.

As is known, for the dual space  $L_1(\Omega, P)^*$  we have
$L_1(\Omega, P)^* = L_\infty (\Omega, P)$ (where $L_\infty (\Omega, P)$ is the Banach space of equivalence classes of essentially bounded functions with  essup--norm). In this case elements  $x^* \in L_\infty (\Omega, P)$ are identified with functionals  (elements of $L_1(\Omega, P)^*$) by means of coupling
\begin{equation}
\label{e.e61}
<x^*,u> = \int_\Omega u\, x^* \,{d\,P}, \ \ \ u\in L_1(\Omega, P).
\end{equation}

Let us consider now the cone   $L_{1+}$  \eqref{e.0'} of nonnegative functions in  $L_1(\Omega, P)$. By  $L_{1+}^*\subset L_1(\Omega, P)^* = L_\infty (\Omega, P)$ we denote the cone of nonnegative functionals on $L_{1+}$, i.e.
\begin{equation}
\label{e.7'''}
L_{1+}^*:= \{{x^*} \in L_\infty (\Omega, P) : \, <x^*,u>\,  \ge 0 \  \text{for every} \ u\in L_{1+}\}.
\end{equation}
Evidently,  $L_{1+}^*$ coincides with the cone  $L_{\infty +}$ of nonnegative functions from  $L_\infty (\Omega, P)$, i.e.
\begin{equation}
\label{e.71'''}
L_{1+}^*= L_{\infty +}:= \{x^*\in  L_\infty (\Omega, P), \ x^*\ge 0\}.
\end{equation}
We denote by   $\tilde{L}_{\infty +}$ the cone
 \begin{equation}
\label{e.72'''}
\tilde{L}_{\infty +}:= \{x^*\in  L^\infty (\Omega, P), \ x^* >_{\text{a.e.}} 0\}.
\end{equation}
By  $L^\perp \subset L_1(\Omega, P)^*$ we denote the subspace of functionals annihilating on  $L$ ($L$ is the subspace  \eqref{e.8} generated by the vectors   $Y^i, \ i=1,\ldots,d$).
\smallskip

As is shown in \cite{ZL} one can rewrite the mentioned fundamental theorem of asset pricing in the following way.
\begin{Tm}
\label{t.4}{\rm [fundamental theorem of asset pricing: geometric formulation: \cite{ZL}, Theorem~1.3]}

 For the objects described above the following two conditions are equivalent:

{\rm 1)} $L\cap L_{1+} = \{0\}$ {\rm(}= absence of arbitrage{\rm)};

{\rm 2)} $L^\perp \cap \tilde{L}_{\infty +} \neq \emptyset$ {\rm(}= existence of a martingale measure{\rm)}.
\end{Tm}

In \cite{ZL}, in  particular, a certain  'refined version' of this result is proven. Namely, let us consider the cone
\begin{equation}
\label{e.72'''1}
\overset{\circ}{L}_{\infty +}= \{x^*\in  L_{\infty +},  \ x^* \ {\text{essentially separated from zero}} \ 0\}.
\end{equation}
Clearly,  $\overset{\circ}{L}_{\infty +}\subset \tilde{L}_{\infty +} $ \  and \  $\overset{\circ}{L}_{\infty +}$ is nothing else as the interior of the cone  $L_{\infty +}$ \eqref{e.71'''}.
\begin{Tm}
\label{t.4'}{\rm [fundamental theorem of asset pricing: 'refined version'; \cite{ZL}, Theorem~1.3]}

For the objects described above the following two conditions are equivalent:

{\rm 1)} $L\cap L_{1+} = \{0\}$ {\rm(}= absence of arbitrage{\rm)};

{\rm 2)} $L^\perp \cap \overset{\circ}{L}_{\infty +} \neq \emptyset$ {\rm(}= existence of a martingale measure   {\rm(}refined condition{\rm)}{\rm)}.
\end{Tm}

Moreover, the main goal of \cite{ZL} was to uncover  a general geometric nature of arbitrage freeness type phenomena.
The corresponding geometric picture is given by the next
\begin{Tm}
\label{t.'4''} {\rm [arbitrage free markets geometry; \cite{ZL}, Theorem~2.4]} Let  $E$ be a Banach space,  $K\subset E$ be a closed cone such that
$\overset{\circ}{K^*} \neq \emptyset$ and  $L\subset E$ be a closed linear reflexive subspace  $L=  L^{**}$.
For the objects mentioned above the following two conditions are equivalent:

{\rm 1)} $L\cap K = \{0\}$ {\rm(}= absence of arbitrage{\rm)};

{\rm 2)} $L^\perp \cap \overset{\circ}{K^*} \neq \emptyset$ {\rm(}= existence of a martingale measure{\rm)};

{\rm 3)}  $L^\perp + \overset{\circ}{K^*} = E^*$, where $E^*$ is the dual space to $E$.
\end{Tm}

As we have noted in Introduction,  once one relaxes essentially conditions on $K$ (for example, do not presupposes satisfaction of the property $\overset{\circ}{K^*} \neq \emptyset$) martingality somehow disappears from arbitrage freeness criterium: in a general situation there is no condition like condition 2) in Theorem~\ref{t.'4''} and only a certain analog of condition 3) takes place (see \cite{ZL}, Theorem~4.1).

The main results of the article are presented in the next section, where we undertake
 the analysis of  relaxation of conditions on  $K$ without loosing the martingality nature of arbitrage free markets.
In addition we will find out that in a number of situations Theorems~\ref{t.4},~\ref{t.4'}, and~\ref{t.'4''} can be strengthened in a way.

\section{Fundamental theorem of asset pricing: a strengthened version}\label{s.2'}

To formulate the results we need to introduce a number of objects.

 Let  $E$ be a Banach space and  $K\subset E$ be a certain cone.
 Recall that  a \emph{cone} in a vector space is a set  $K$ possessing the following two properties:

1) $K$ is a convex set;

2) for every  $x\in K$ and any  $0<\lambda \in \mathbb R$ one has  $\lambda x\in K$.
\smallskip

 As usually by   ${K}^*$ we denote the cone of nonnegative functionals on  $K$, i.e.
\begin{equation}
\label{e.12}
{K}^*:= \{{x^*} \in E^* : \, <x^*,u>\,  \ge 0 \  \text{for any} \ u\in K\},
\end{equation}
here  $E^*$ is the space dual to  $E$.
\smallskip

Let  $K=\overline{K}$ be a closed cone. A cone  ${\cal K}\subset K^*$ is called  \emph{total} if it possesses the following property: let  $u\in E$ and for every  $x^*\in \cal K$ one has  $<x^*, u> \, \geq 0$, then  $u\in K$.  The set of all total cones in  $K^*$ will be denoted by ${\bf K}^*_{tot}$.
\smallskip

{\bf Examples}  1. Clearly the cone  $K^*$ itself is total.

 2. If  $K^*$ has a nonempty  interior  $\overset{\circ}{K^*} \neq \emptyset$, then  $\overset{\circ}{K^*}$  is a total cone. Indeed, in this case we have   $\overline{\overset{\circ}{K^*}} = K^*$. If  $u\notin K$, then there exists a functional
$x^* \in K^*$, such that  $<x^*, u> \ < 0$. Therefore for functionals  $x^{*\prime} \in \overset{\circ}{K^*}$ that are sufficiently close to $x^*$  one has  $<x^{*\prime} , u> \ < 0$ \ as well.

 3. In particular, if  ${L}_{\infty +}$ is the cone  \eqref{e.71'''} of nonnegative functions in  $L_\infty (\Omega, P)$, which is interpreted as the cone  $L_{1+}^*$ of nonnegative functionals on the cone  $L_{1+}$  of nonegative functions in  $L_1(\Omega, P)$ \, \eqref{e.0'}, then the cone
\begin{equation}
\label{e.72'''1}
\overset{\circ}{L}_{\infty +}= \{Q\in  L_{1+}^*(= {L}_{\infty +}),  \ Q \ {\text{is essentially separated from zero}}\}.
\end{equation}
is a total cone.

 4. For the cone  $L_{1+}^*= {L}_{\infty +}$ \eqref{e.71'''} let us consider one more cone   $\tilde{L}_{\infty +}$ \eqref{e.72'''}. Clearly,
 \begin{equation}
\label{e.74'''}
<Q,f> \, \ge 0\ \, \text{for every} \ Q\in \tilde{L}_{\infty +} \ \Leftrightarrow \ f\in L_{1+}\,,
\end{equation}
that is  $\tilde{L}_{\infty +}$
  is a total cone as well.

5. Let again  $L_{1+}^*= {L}_{\infty +}$ be the cone   \eqref{e.71'''}, and  $\cal K$ be the cone of simple
positive functions having finite sets of values, i.e. finite sums of the form
$$
f = \sum_i c_i \chi (A_i),
$$
where $\Omega \supset A_i $ are measurable sets,  $\coprod_i A_i =\Omega$ and  $c_i >0$.

\noindent Clearly  $\cal K$ is a total cone.
\medskip

A cone  ${\cal K} \subset K^*$ is said to be  \emph{nonannihilating} if it possesses the following property:  for every  $x^* \in\cal K$ and any  $0 \neq u \in K$ one has  $<x^* , u> \, > 0$. The set of all nonannihilating cones in  $K^*$ is denoted by ${\bf K}^*_{nan}$.

\smallskip

{\bf Examples}. \
 6. If  $K^*$ has a nonempty interior  $\overset{\circ}{K^*} \neq \emptyset$, then  $\overset{\circ}{K^*}$  is a nonannihilating cone. Indeed, suppose that $x^*\in \overset{\circ}{K^*}$ and   $<x^*, u> \, = 0$. In this case let us take any functional $y^*\in E^*$ such that  $<y^*, u> = \alpha <0$. Since  $x^*\in \overset{\circ}{K^*}$ it follows that for sufficiently small positive $\varepsilon$ one has  $x^*+ \varepsilon y^*\in \overset{\circ}{K^*}$ and therefore  $<x^*+ \varepsilon y^*, u> = \varepsilon\alpha <0$. Thus we arrived at a contradiction.

7. In particular, if  $L_{1+}^*= {L}_{\infty +}$ is the cone \eqref{e.71'''} then the cone  $\overset{\circ}{L}_{\infty +}$  \ \eqref{e.72'''1}
is nonannihilating.

8. For the same cone $L_{1+}^*= {L}_{\infty +}$  \eqref{e.71'''} the cone  $\tilde{L}_{\infty +}$ \eqref{e.72'''} is nonannihilating as well.

 9. Let again $L_{1+}^*= {L}_{\infty +}$ be the cone  \eqref{e.71'''} and  $\cal K$  be the cone of simple
positive functions having finite sets of values. Clearly,   $\cal K$ is a nonannihilating cone.

10. Unfortunately not for every cone  $K$ one has  ${\bf K}^*_{nan}\neq \emptyset$. Let, for example,  $K=L$ be a linear subspace. In this situation  $L^* = L^\perp$. Thus, for every  $x^*\in L^*$ and for any  $u\in L$ we have  $<x^*, u> =0$.
\medskip

\medskip
The first essential result in description of arbitrage free markets is the following observation.

\begin{La}
\label{l.14}
Let  $E$ be a Banach space and  $K, L \subset E$, where  $K$ is a cone and   $L$ is a linear subspace. For every nonannihilating  ${\cal K} \in {\bf K}^*_{nan}$ the following relation is true
\begin{equation}
\label{e.16}
L^\perp \cap {\cal K} \neq \emptyset \  \Rightarrow \ L\cap K =\{0\}.
\end{equation}
\end{La}

\emph{Proof}. \  Implication   \eqref{e.16} is equivalent to the implication
 $$
L\cap K \neq \{0\} \  \Rightarrow  \ L^\perp \cap {\cal K} = \emptyset.
 $$
Therefore we prove the latter one.

Let
$$
0\neq u_0 \in L\cap K .
$$
Consider the functional  $\overline{u}_0 \in  E^{**}$ given by relation
$$
\overline{u}_0 (x^*) := \, <x^*,{u}_0>,  \ \, x^*\in E^*.
$$
To prove the equality  $L^\perp \cap {\cal K}  = \emptyset$ it is enough to check that the functional  $\overline{u}_0$  strictly  separates $L^\perp$ and ${\cal K}$, that is
\begin{equation}
\label{e.e1}
\text{for every} \ x^*\in L^\perp \ <x^*,{u}_0> =0,
\end{equation}
\begin{equation}
\label{e.e2}
\text{for every} \ x^*\in {\cal K} \ <x^*,{u}_0>\,  >0.
\end{equation}
Equality  \eqref{e.e1} follows from the fact that  ${u}_0\in L$, and inequality  \eqref{e.e2} is true since  ${\cal K} \in {\bf K}^*_{nan}$. The lemma is proved.
\medskip

The next observation is presented by
\begin{La}
\label{l.15}
Let  $E$ be a Banach space and $K, L \subset E$, where $K= \overline{K}$ is a closed cone, and   $L$ is a finite-dimensional linear subspace.  For every total cone  ${\cal K} \in {\bf K}^*_{tot}$ the following implication is true \begin{equation}
\label{e.17}
L\cap K =\{0\} \  \Rightarrow \   L^\perp \cap {\cal K} \neq \emptyset .
\end{equation}
\end{La}

\emph{Proof}. \
Let us introduce the set
\begin{equation}
\label{e.51'}
C:= \left\{ x^*|_L, \  x^*\in {\cal K}\right\} \subset L^*,
\end{equation}
that is the set of restrictions of functionals  ${\cal K}$ onto $L$ (here  $L^*$ is the dual space to $L$,
that is the set of linear continuous functionals on  $L$). By assumption  $L$ is  finite-dimensional. Therefore,   $L\cong L^*$.

Evidently,
\begin{equation}
\label{e.52'}
L^\perp \cap {\cal K} \neq \{\emptyset\} \ \Leftrightarrow \ C \ni 0,
\end{equation}
where $0\in L^*$.

Relation  \eqref{e.52'} shows that to prove  \eqref{e.17} it is enough to verify the following implication
\begin{equation}
\label{e.53'}
0 \notin C \ \Rightarrow \  L \cap {K} \neq \{0\}.
\end{equation}
Therefore, hereafter we prove  \eqref{e.53'}.

Since  ${\cal K}$ is a cone we have that    $C$ is a cone in  $L^*$.

As  $C$ is a cone and  $0 \notin C $ it follows (by separation theorem) that there exists  $f_0 \in L$, such that

i) for every  $x^*\in {\cal K} \ <x^*,f_0> \, \ge 0$;

ii) there exists  $x^*_0 \in {\cal K}$, such that  $<x^*_0, f_0> \, > 0$.
\smallskip

Since  ${\cal K} \in {\bf K}^*_{tot}$ condition  i) implies  $f_0 \in {K}$; and condition  ii) implies  $f_0 \neq 0$.
So, \eqref{e.53'} is verified and the proof is complete.
\smallskip

Combining Lemmas~\ref{l.14} and \ref{l.15} we obtain
\begin{Tm}
\label{str-principal-t} {\rm [fundamental theorem of asset pricing: a strengthened version]} \ Let  $E$ be a Banach space and  $K, L \subset E$, where  $K= \overline{K}$ is a closed cone, and $L$ is a finite-dimensional linear subspace. Suppose that  ${\bf K}^*_{tot} \cap {\bf K}^*_{nan} \neq \emptyset$. Then for every cone
${\cal K} \in {\bf K}^*_{tot} \cap {\bf K}^*_{nan}$ the following to conditions are equivalent:

{\rm 1)} $L\cap K =\{0\}$  {\rm(}= absence of arbitrage{\rm)};

{\rm 2)} $L^\perp \cap {\cal K} \neq \emptyset$ {\rm(}= existence of a martingale measure {\rm(}a strengthened condition{\rm)}{\rm)}.
\end{Tm}

Note that the above presented Examples 2 and 6 show that if the cone  $K^*$ has a nonempty interior  $\overset{\circ}{K^*} \neq \emptyset$, then  $\overset{\circ}{K^*} \in {\bf K}^*_{tot}\cap{\bf K}^*_{nan}$. Thus Theorem~\ref{str-principal-t} for a finite-dimensional subspace $L$, in particular, implies equivalence of 1) and 2) in Theorem~\ref{t.'4''}.
\smallskip

The next figure illustrates the difference between Theorems~\ref{t.4},~\ref{t.4'} and~\ref{str-principal-t}  by means of the cone   $\cal K$ from Examples~5~and~9. Note that in a general situation one has the following relations for the cones $\tilde{L}_{\infty +}$ (Theorem~\ref{t.4}), $\overset{\circ}{L}_{\infty +}$ (Theorem~\ref{t.4'}) and $\cal K$ (Examples~5~and~9):
$$
\tilde{L}_{\infty +} \, \underset{\neq}{\supset} \, \overset{\circ}{L}_{\infty +} \, \underset{\neq}{\supset} \,  {\cal K}
$$
and therefore Theorem~\ref{str-principal-t} is the strongest assertion among the mentioned ones.

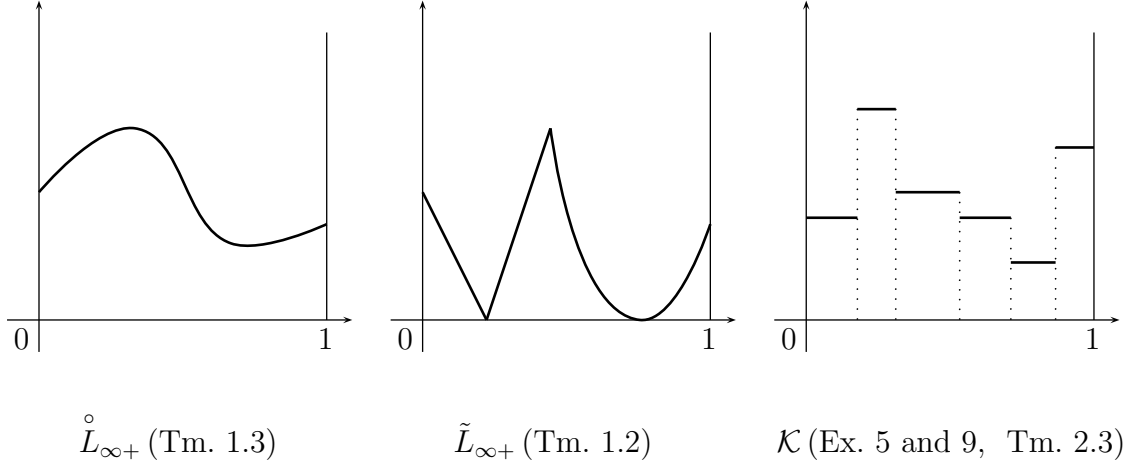
\psset{linewidth=0.5pt, unit=8.5mm}
\begin{center}
\begin{pspicture}(18.0,6.0)
 \psline{->}(0.5,1.0)(5.9,1.0)  \psline{->}(1.0,0.5)(1.0,6.0)  \psline{-}(5.5,1.0)(5.5,5.5)
 \uput[0](0.4,0.7){$0$}  \uput[0](5.15,0.7){$1$}
 \pscurve[linewidth=1.0pt](1.0,3.0)(2.5,4.0)(4.0,2.2)(5.5,2.5)
 \psline{->}(6.5,1.0)(11.9,1.0)  \psline{->}(7.0,0.5)(7.0,6.0)  \psline{-}(11.5,1.0)(11.5,5.5)
 \uput[0](6.4,0.7){$0$}  \uput[0](11.15,0.7){$1$}
 \psline[linewidth=1.0pt](7.0,3.0)(8.0,1.0)
 \psline[linewidth=1.0pt](8.0,1.0)(9.0,4.0)
 \pscurve[linewidth=1.0pt](9.0,4.0)(10.4,1.0)(11.5,2.5)
 \psline{->}(12.5,1.0)(17.9,1.0)  \psline{->}(13.0,0.5)(13.0,6.0)  \psline{-}(17.5,1.0)(17.5,5.5)
 \uput[0](12.4,0.7){$0$}  \uput[0](17.15,0.7){$1$}
 \psline[linewidth=1.0pt](13.0,2.6)(13.8,2.6)
 \psline[linewidth=1.0pt](13.8,4.3)(14.4,4.3)
 \psline[linewidth=1.0pt](14.4,3.0)(15.4,3.0)
 \psline[linewidth=1.0pt](15.4,2.6)(16.2,2.6)
 \psline[linewidth=1.0pt](16.2,1.9)(16.9,1.9)
 \psline[linewidth=1.0pt](16.9,3.7)(17.5,3.7)
 \psline[linewidth=0.7pt,linestyle=dotted](13.8,1.0)(13.8,4.3)
 \psline[linewidth=0.7pt,linestyle=dotted](14.4,1.0)(14.4,4.3)
 \psline[linewidth=0.7pt,linestyle=dotted](15.4,1.0)(15.4,3.0)
 \psline[linewidth=0.7pt,linestyle=dotted](16.2,1.0)(16.2,2.6)
 \psline[linewidth=0.7pt,linestyle=dotted](16.9,1.0)(16.9,3.7)
\end{pspicture}
\end{center}
$$
\ \ \ \ \ \ \ \ \ \ \overset{\circ}{L}_{\infty +} \, (\text{Tm}.~\ref{t.4'})\ \ \ \ \   \ \ \ \ \ \ \ \ \ \ \ \ \tilde{L}_{\infty +} \, (\text{Tm}.~\ref{t.4})   \ \ \ \ \ \ \ \ \ \ \ \ {\cal K} \,
(\text{Ex.~5~and~9, \ Tm.~\ref{str-principal-t}})
$$

  One has to emphasize that  property $\overset{\circ}{K^*} \neq \emptyset$ is rather special (see in this connection a discussion of this property in \cite{ZL}, Section~3). This property takes place for the cone  $L_{1+}^*= {L}_{\infty +}$  \eqref{e.71'''} (and namely this cone is exploited in Theorem~\ref{t.4}). On the other hand, if, for example, one considers  the analogous cones  ${L}_{p+}$  of nonnegative functions in the spaces  $L_p (\Omega, P), \ 1\le p <\infty$, then here one has  $\overset{\circ}{L}_{p+}= \emptyset$. In addition for  $1\le p<\infty$  we have  ${L}_{p+}^* = {L}_{q+}$, where  $\frac{1}{q} + \frac{1}{p} = 1$ and we assume $\frac{1}{\infty} =0$. Let us note, however, that for every cone  $K:={L}_{p+}, \ 1\le p<\infty$ one again has  ${\bf K}^*_{tot}\cap{\bf K}^*_{nan} \neq \emptyset$. Indeed, for example, the following three  cones are elements of this set

a) ${\cal K}_1 = \{Q\in  L_{q+}, \ Q\ge_{\text{a.e.}} 0\}$;

b) ${\cal K}_2 = \{Q\in {\cal K}_1, \ Q \ {\text{is essentially separated from zero}}\}$;

c) ${\cal K}_3$ is the cone of simple
positive functions having finite sets of values.
\smallskip

Therefore, Theorem~\ref{str-principal-t}, in particular, implies the next result, which is reasonable to consider as a  strengthened version  of  the fundamental theorem of asset pricing for $p$-summable markets.

\begin{Tm} {\rm [fundamental theorem of asset pricing for $p$-summable markets]}
\label{L-p}  \ Let  $K:={L}_{p+}$  be the cone of nonnegative functions in  {$L_p (\Omega, P), \ 1 \le p <\infty$} \ and \ $L\subset L_p (\Omega, P)$  be a finite-dimensional linear subspace. Then for every of the cones ${\cal K}_i, \ i=1,2,3$ mentioned above the following two conditions are equivalent:

{\rm 1)} $L\cap K =\{0\}$ {\rm(}= absence of arbitrage{\rm)};

{\rm 2)} $L^\perp \cap {\cal K}_i \neq \emptyset$ {\rm(}= existence of a martingale measure{\rm)}.
\end{Tm}

To finish the article we give two comments on the subspace finite dimensionality condition. First of all note that in the situation $E={\mathbb R}^n$  the set ${\bf K}^*_{tot}\cap{\bf K}^*_{nan}$ reduces to a single element.
\begin{Tm}
\label{t.finite-dim}
If $K\subset {\mathbb R}^n$ is a closed cone such that $K\cap (-K) =0$ then
$\overset{\circ}{K^*}\neq \emptyset$ and
$$
{\bf K}^*_{tot}\cap{\bf K}^*_{nan} = \{\overset{\circ}{K^*}\} \neq \emptyset \,.
$$
\end{Tm}
\emph{Proof}. \ Note that under the conditions of theorem $K$ is a plasterable cone (see  \cite{ZL}, Example~3.3.) and we have $\overset{\circ}{K^*} \neq \emptyset$. Thus (see Examples 2 and 6)
$$
\overset{\circ}{K^*} \in {\bf K}^*_{tot}\cap{\bf K}^*_{nan}\,,
$$
and, in particular,  ${\bf K}^*_{tot}\cap{\bf K}^*_{nan}\neq \emptyset$.

Now let us consider any element ${\cal K} \in {\bf K}^*_{tot}\cap{\bf K}^*_{nan}$. We have to prove that ${\cal K}= \overset{\circ}{K^*}$.

Since $K$ is plasterable \cite{ZL}, Theorem~3.1., 5) implies that
$$
F:=\overline{co}\left( K\cap \{u :\|u\|=1\}\right) \not\ni 0\,,
$$
where  $co (A)$ is the convex hull of  $A$.

As ${\cal K} \in {\bf K}^*_{nan}$ one has that
\begin{equation}
\label{ed.1}
\langle x^*, u\rangle >0 \ \ \text{for every} \ x^* \in {\cal K} \ \text{and every} \ u\in F.
\end{equation}
Since $F$ is compact  \eqref{ed.1} implies that for every $x^* \in {\cal K}$ there exists  a constant $c>0$ such that
\begin{equation}
\label{ed.2}
\langle x^*, u\rangle >c \ \ \text{for every}  \ u\in F.
\end{equation}
Therefore
$$
\langle x^*, u\rangle >c \Vert u\Vert \ \ \text{for every}  \ 0\neq u\in K.
$$
Which means that $x^* \in \overset{\circ}{K^*}$ (by \cite{ZL}, Theorem~3.1., 3)). So we have proved the inclusion
$$
{\cal K} \subset \overset{\circ}{K^*}.
$$

Now let us prove the opposite inclusion.

First of all note that
\begin{equation}
\label{ed.3}
\overline{\cal K} = K^*.
\end{equation}
The proof goes by contradiction. Suppose that
$$
\overline{\cal K} \neq K^*.
$$
It means that there exists $y^*\in K^*$ such that $y^* \not\in \overline{\cal K}$. By separation theorem we conclude that there
exists $u\in {\mathbb R}^n$ such that
\begin{equation}
\label{ed.4}
\langle x^*, u\rangle \geq 0 \ \ \text{for every} \ x^* \in {\cal K},
\end{equation}
and
\begin{equation}
\label{ed.5}
\langle y^*, u\rangle < 0\,.
\end{equation}
 Since ${\cal K} \in {\bf K}^*_{tot}$ it follows that \eqref{ed.4} implies $u\in K$ which contradicts \eqref{ed.5}. Thus \eqref{ed.3} is proved.

 Now take any $y^* \in \overset{\circ}{K^*}$. By Caratheodory's Theorem there exists $\{y^*_1, \ldots , y^*_{n+1}\}\subset K^*$ such that
\begin{equation}
\label{ed.6}
y^* \in \overset{\circ}{co(\{y^*_1, \ldots , y^*_{n+1}\})}
\end{equation}
Bearing in mind \eqref{ed.3} and taking $\{x^*_1, \ldots , x^*_{n+1}\}\subset {\cal K}$ in such a way that
each $x^*_i, \ i={1, \ldots, n+1}$ is sufficiently close to the corresponding $y^*_i$ we obtain from \eqref{ed.6} that
$$
y^* \in \overset{\circ}{co(\{x^*_1, \ldots , x^*_{n+1}\})}\,.
$$
Thus $\overset{\circ}{K^*} \subset{\cal K}$ and the proof is finished.
\smallskip

As a corollary of the theorem just proved we conclude that  in the situation under consideration Theorem~\ref{str-principal-t} reduces to
\begin{Tm}
\label{t.finite}
Consider the space  ${\mathbb R}^n$.
Let $ K\subset {\mathbb R}^n$ be a closed cone such that \ $K\cap (-K)= \{0\}$.  For a linear subspace  $L\subset {\mathbb R}^n$ the following two conditions are equivalent:
\smallskip

{\rm 1)} $L\cap K = \{0\}$,

{\rm 2)} $L^\perp \cap \overset{\circ}{K^*} \neq \emptyset$, where   $L^\perp$ the orthogonal
complement to  $L$.
\end{Tm}
Our final observation shows  that in general finite dimensionality condition for the subspace  $L$ is essential in formulation of the strengthened version of the fundamental theorem of asset pricing (if  the dimension of $L$ is infinite then not for every  ${\cal K} \in {\bf K}^*_{tot} \cap {\bf K}^*_{nan}$ properties  1) and  2) of Theorem~\ref{str-principal-t} are equivalent).

{\bf Example}.  Let $E = l_1 = \{ (\xi_1, \xi_2, \dots ): \ \sum_i |\xi_i|<\infty\}$ and  let \ $K\subset l_1$ be the cone of nonnegative sequences and  $L$ be the subspace generated by  vectors of the form  \ $e_{2n}- \frac{1}{2n}e_{2n-1}, \ n=1,2, \dots$, \ where $e_k, k=1,2,\dots$ is the canonical base in  $l_1$.  Evidently $L$ is nothing else than the space of  vectors of the form  $\sum_{n=1}^\infty \xi_n (e_{2n}- \frac{1}{2n}e_{2n-1}):  \sum_{n=1}^\infty |\xi_n | <\infty$. Clearly
$$
K\cap L= \{0\}.
$$
The dual space to  $l_1$ is the space  $l_\infty$ of bounded sequences and the action of an element   $f = (\nu_1, \nu_2, \dots )\in l_\infty$ on $x= (\xi_1, \xi_2, \dots )\in l_1$ is given by the coupling
$f(x)= \sum_i \xi_i\,\nu_i$.
In this example   $K^*\subset l_\infty$ is the cone of nonnegative bounded sequences.
 Let ${\cal K}=\overset{\circ}{K^*}$ be the set of  positive sequences separated from zero
 (clearly  $\overset{\circ}{K^*} \in {\bf K}^*_{tot} \cap {\bf K}^*_{nan}$).
Evidently, if  $f = (\nu_1, \nu_2, \dots )\in L^\perp$ then    $\nu_{2n-1}= \frac{1}{2n}\nu_{2n}$.  This along with the boundness of the sequence  $(\nu_{2n})$ implies  $\nu_{2n-1}\to 0$. Therefore   $f\notin {\cal K}$, that is
$$
L^\perp \cap {\cal K} = \emptyset .
$$
Thus in this example ${\bf K}^*_{tot} \cap {\bf K}^*_{nan}\neq \emptyset$ and here condition  1) of Theorem~\ref{str-principal-t} is satisfied while condition  2) is not satisfied.

\bigskip

 To finish the article we note  that the material presented naturally causes the problem: can we obtain a general description of  the cones $K$ possessing the property
${\bf K}^*_{tot} \cap {\bf K}^*_{nan}\neq \emptyset$?


\def\refname{
\bigskip

{\large
\noindent Department of Mechanics and Mathematics, Belarus State University,\\
PR. Nezavisimosti,4, 220050, Minsk, Belarus\\
 Institute of Mathematics,  University  of Bialystok,\\
 ul. Akademicka 2, PL-15-267  Bialystok, Poland\\}
 {\it e-mail}:\ {lebedev@bsu.by}
\medskip

{\large
\noindent Department of Mechanics and Mathematics, Belarus State University,\\
PR. Nezavisimosti,4, 220050, Minsk, Belarus\\}
 {\it e-mail}:\ {zabreiko@mail.ru}
\end{document}